\documentstyle[prl,multicol,epsf,aps]{revtex}
\draft

\title{Anomalous L\'evy decoherence} 
\author{Eric Lutz} 
\address{D\'epartement de Physique Th\'eorique, Universit\'e de Gen\`eve, 24, 
quai Ernest Ansermet, 
1211 Gen\`eve 4, Switzerland} 

\date{\today}

\def\openone{\leavevmode\hbox{\small1\kern-3.3pt\normalsize1}}

\newcommand{\la}{\langle}
\newcommand{\ra}{\rangle}
\newcommand{\be}{\begin{equation}}
\newcommand{\ee}{\end{equation}}
\newcommand{\bea}{\begin{eqnarray}}
\newcommand{\eea}{\end{eqnarray}}

\newcommand{\SC}{\scriptscriptstyle}

\begin{document}

\twocolumn[\hsize\textwidth\columnwidth\hsize\csname@twocolumnfalse\endcsname
\maketitle

\begin{abstract}
We investigate the decoherence of a small quantum system weakly coupled to a complex, 
chaotic environment when the dynamics is not Gaussian but L\'evy anomalous. By studying 
the time dependence of the linear entropy and the damping of  the interference  of two 
Gaussian wave packets in the Wigner representation, we show that   the decoherence time for a quantum L\'evy stable process is 
{\it always} smaller than for  Gaussian diffusion. 
\end{abstract}
\pacs{PACS numbers:  05.40.Fb, 03.65.Yz, 05.40.-a}
 \vskip.5pc]
First a subject of mathematical investigation \cite{lev37,doo42},  L\'evy stable processes have  nowadays
fully entered the realm of physics \cite{shl94,pek98}.  It is well known that  usual Brownian motion is a Gaussian stochastic process \cite{ris89}. This is a consequence of  the Gaussian law of large numbers, or central--limit theorem. In the 1930s, the  central--limit theorem has been extended  by L\'evy and others to random variables with infinite second moment.  More precisely, the probability distribution of these random variables has the form,
\be
\label{eq1}
{\cal L}^{\SC{C}}_\alpha(x)=\frac{1}{2\pi}\int_{-\infty}^\infty e^{ikx-C |k|^\alpha}dk\ , \hspace{.5cm} 
0<\alpha \leq 2 \ .
\ee
A (symmetric) L\'evy  stable distribution as given by  (\ref{eq1}) can be regarded as an extension of the Gaussian distribution to which it reduces when $\alpha\!=\!2$. It should be stressed, however, that the Gaussian distribution is the only stable law having a finite variance.
A L\'evy process may then be defined as a non--Gaussian  
generalization  of Brownian motion obeying L\'evy stable statistics. 
The hallmark of this form of anomalous diffusion is the presence of long jumps 
(the so--called ``L\'evy flights'') which are due to the asymptotic 
power--law tail of a stable distribution, ${\cal L}^{\SC{C}}_\alpha(x) \sim 1/|x|^{\alpha+1}$.
As a consequence, the mean--square displacement and the mean 
kinetic energy of a L\'evy particle are divergent \cite{jes99}. L\'evy dynamics often results 
from the interaction with a complex, non-homogeneous environment. Examples are  porous or disordered media \cite{bou90} or  chaotic heat baths \cite{kus99}. 

The first experimental observation of  a L\'evy stable process has been reported 
 ten years ago in micelle systems \cite{ott90}. Since then, L\'evy--like diffusion has for 
instance  been seen 
in two--dimensional rotating flows \cite{sol93}, in porous glasses \cite{sta95} and, more recently,  in ion lattices \cite{kat97}
  and in subrecoil
 laser cooling \cite{sau99}. The ion lattice experiment is of particular interest since it reports a  direct  measurement of the divergence of the mean kinetic energy of the particle.  
Moreover, it is   worthwhile to notice that  experiments have now started 
to study anomalous L\'evy dynamics of microscopic systems (atoms, ions) where quantum effects are likely to play a
role. This motivated us  to examine  the decoherence
of   quantum L\'evy stable motion.  

Decoherence  manifests itself in the dynamical destruction of quantum interferences as a result of the interaction with the environment \cite{zur81}.  This progressive  transition from quantum to 
classical  has been experimentally observed in high Q microwave cavities \cite{bru96}. Theoretically,  decoherence   has been  intensively investigated in the case of normal 
(Gaussian) diffusion \cite{giu96,paz99,zur01} but not so far for anomalous L\'evy diffusion. 

A simple way to evaluate the decoherence time $\tau^\alpha_D$ is to compute the linear entropy $S_\alpha(t)= 1- \mbox{tr} \rho^2(t)$, which gives a measure of  the purity of the system \cite{zur93,kim96}: The linear entropy is equal to zero for a pure state and increases to one when the state of the system evolves to a mixture. Here $\rho(t)$ is the reduced density operator of the system. Alternatively, one can consider  the superposition of two Gaussian wave packets in the Wigner representation and look at the damping of the interference fringes \cite{joo85}.  In the following we study the time dependence of both the linear entropy and of 
the interference term of the Wigner function. We put  a special emphasis on the value $\alpha=1$ where analytical results can be obtained. What we  find is that L\'evy 
decoherence is {\it always}  faster than Gaussian decoherence for any value of the stability index $\alpha$.  

Starting point of our analysis is a master equation for a  system coupled to a chaotic environment  derived by Kusnezov {\it et al.}  from a microscopic  random--matrix model \cite{kus99}. In this approach the coupling to the non--homogeneous
  background is characterized  by a spreading width $\Gamma^\downarrow$ and a spatial correlation function $G(x/x_0)$ with 
an environment characteristic  correlation length $x_0$ (see Ref.~\cite{kus99} for details). In the limit of high (actually infinite) 
temperature, the quantum master equation for a  free particle of mass $M$
 is given by
\bea
\label{eq2}
\frac{\partial}{\partial t}\rho(x,y,t)& =& -\frac{i\hbar}{2M} \Big[\frac{\partial^2}{\partial x^2} -\frac{\partial^2}{\partial y^2}\Big]  \rho(x,y,t) \nonumber \\
&-&\frac{\Gamma^\downarrow}{\hbar}\Big [G\Big(\frac{x-y}{x_0}\Big)-1\Big] \rho(x,y,t)\ . 
\eea
When the  environment correlation length  $x_0$ is much larger than the separation $ \Delta x=|x-y|$,  the correlator in Eq.~(\ref{eq2}) can be expanded as\cite{rem1}
\be
\label{eq3}
  G\left(\frac{\Delta x}{x_0}\right ) = 1- \left|\frac{\Delta x}{x_0}\right|^\alpha\ .
\ee
 The case $\alpha\!=\!2$ corresponds to Gaussian  diffusion, while for a  faster spatial decorrelation, $0<\!\alpha\!<2$, the diffusion is L\'evy anomalous  \cite{kus99,lut01}.  The (generalized)  diffusion coefficient is further defined as $D_\alpha= \Gamma^\downarrow/(\hbar x_0^\alpha)$. It is also worthwhile to emphasize  that it is the last term in the master equation (\ref{eq2}), which results from the accumulation of  phase fluctuations of the environment,  which is mainly responsible for the  decoherence effect \cite{zur81}. We already know  that the alteration of this noise term  when passing  from Gauss to L\'evy statistics has a dramatic effect on the diffusion properties of the system, it is thus of high interest to examine the effect of the modification of this term on decoherence. 

We begin by solving  the master equation  (\ref{eq2}) for $\rho(t)$. For this purpose it is convenient to use the $(k,\Delta)$ representation of the density matrix \cite{unr89},
\be
\label{eq4}
\rho(k,\Delta,t)= \int _{-\infty}^\infty dq \,e^{ikq} \, \rho(q+\frac{\Delta}{2},q-\frac{\Delta}{2},t) \ ,
\ee
instead of the coordinate representation $\rho(x,y,t)$. Here, we have introduced the center of mass and relative coordinates, $q=(x+y)/2$ and $\Delta= x-y$. In this representation, the master equation   reads
\be
\label{eq5}
\frac{\partial}{\partial t} \rho(k,\Delta,t)= \Big\{\frac{k}{M} \frac{\partial}{\partial \Delta} -D_\alpha |\Delta|^\alpha \Big\} \rho(k,\Delta,t) \ .
\ee
This first order partial differential equation can easily be solved, for instance by the method of 
characteristics . We find 
\bea
\label{eq6}
\rho(k,\Delta,t)& =& \rho(k,\Delta+\frac{kt}{M},0) \,\exp\left[-\frac{M D_\alpha}{(1+\alpha)k} \left(\Delta +\frac{kt}{M}\right) \right.\nonumber \\
&&\times \left. \left|\Delta+\frac{kt}{M}\right|^\alpha \right] \exp\left[\frac{M D_\alpha}{(1+\alpha)k}\, \Delta |\Delta|^\alpha \right] \ ,
\eea
where $\rho(k,\Delta,0)$ is the reduced density operator at $t=0$. For 
an initial Gaussian pure state, $\rho(x,y,0)\! =\! \varphi(x) \varphi^*(y)$ with $\varphi(x) = \exp[-x^2/4b^2]/(2\pi b^2)^{1/4}$, it has the form
\be
\label{eq7}
\rho(k,\Delta,0) = \exp\Big[-\frac{k^2b^2}{2} -\frac{\Delta^2}{8b^2} \Big] \ .
\ee
It is further interesting to consider the probability distributions for position and momentum that correspond to the solution (\ref{eq6}). In the limit of long times, we have
\bea
\label{eq8}
\rho(x,x,t) &=& \int_{-\infty}^\infty  dk \,e^{-ikx} \rho(k,\Delta\!=\!0,t) \simeq {\cal L}_\alpha^{u(t)}(x) \ , \\
\label{eq9}
\rho(p,p,t) &=& \int_{-\infty}^\infty  d\Delta \,e^{-ip\Delta} \rho(k\!=\!0,\Delta,t) \simeq {\cal L}_\alpha^{v(t)}(p) \ ,
\eea
with $u(t) = D_\alpha t^{1+\alpha}/(1+\alpha)M^\alpha$ and $v(t)= D_\alpha t$. The result (\ref{eq8}), which holds for $1\!<\!\alpha\!<\!2$, has already been obtained in Ref.~\cite{kus99}. Notice that both the position and the momentum distributions of the system asymptotically approach a L\'evy stable law (in the short time limit they are given by Gaussians). In particular, this means that for $\alpha\!=\!1$, $\rho(p,p,t)$ has the form of a  Lorentzian.   It is worth mentioning that a recent experiment confirmed that the momentum distribution of atoms cooled by selective coherent population trapping is very close to a Lorentzian distribution \cite{sau97}.

\noindent
{\bf Linear entropy.} Let us now  calculate the linear entropy $S_\alpha(t)$ of the system. We  use the expression  (\ref{eq6}) for $\rho(k,\Delta,t)$ together with the initial condition (\ref{eq7}). For short times, we then obtain
\bea
\label{eq10}
S_\alpha(t) &\simeq&1 - \frac{1}{\sqrt{4\pi b^2}} \int d\Delta \,\exp\left[-\frac{\Delta^2}{4b^2}\right] \left(1-2 D_\alpha |\Delta| ^\alpha t \right ) \nonumber \\
&=&\frac{2^{\alpha+1}}{\sqrt{\pi}} \Gamma\left(\frac{1+\alpha}{2}\right)   \frac{\Gamma^\downarrow}{\hbar}\left(\frac{b}{x_0}\right)^\alpha t \ .
\eea
Consequently, the system gets entangled with the environment within  a decoherence time of the order of  $\tau^\alpha_D\sim [(\Gamma^\downarrow/\hbar)(b/x_0)^\alpha] ^{-1}$, where $b$ is the width of the initial wave packet.

\begin{figure}[t]
\centerline{\epsfxsize=7.3cm
\epsfbox{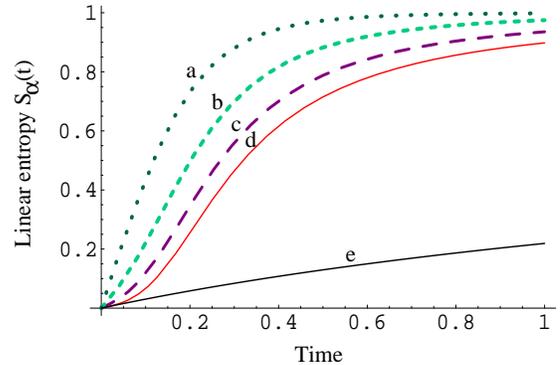}}
\vspace{.1cm}
\caption{Linear entropy $S_\alpha(t)$ as a function of time for different values of the stability index $ \alpha $: The curves $(a)$ to $(d)$ are obtained  with the exact solution (\ref{eq6}) of the master equation for increasing values of $\alpha$, $\alpha =0.5, 1, 1.5, 2$. The curve $(e)$ corresponds to the approximate solution (\ref{eq12}) in the case $\alpha=2$.} 
\label{fig1}
\end{figure}

In order to obtain more tractable  expressions, we will  work in the following  in the  limit $\hbar \rightarrow 0$, where  $D_\alpha \la (\Delta x)^\alpha\ra$ is large \cite{zur01}. In this ``semiclassical'' limit, the time evolution of the density operator is dominated by the last term in  Eq.~(\ref{eq2}). It is then straighforward to write down an approximate solution of the master equation in the form,
\be
\label{eq11}
\rho(x,y,t) = \rho(x,y,0)\, \exp\left[-\frac{\Gamma^\downarrow}{\hbar}\left|\frac{x-y}{x_0}\right|^\alpha t\right]\ .
\ee
We  observe  that the off--diagonal matrix elements  decay exponentially on a time scale $\tau^\alpha_D= [(\Gamma^\downarrow/\hbar)(\Delta x/x_0)^\alpha] ^{-1}$. Note that this  expression of the decoherence time is in agreement with the  previous result  obtained   from the short time behavior of the  linear entropy.
Let us briefly discuss the validity of this approximation. To this end,  it
 is  instructive to write  the approximate solution (\ref{eq11}) of the master
 equation in the $(k,\Delta)$ representation which yields
\be
\label{eq12}
\rho(k,\Delta,t)  =\rho(k,\Delta,0)\exp[-D_\alpha |\Delta|^\alpha t] \ .
\ee
This expression is easily seen to be  a short time approximation of the exact solution (\ref{eq6}) [see also Fig.~(\ref{fig1})].  

\noindent
{\bf Superposition of wave packets.}
We now turn to the study of the effect of decoherence on the interference of two Gaussian wave packets in the Wigner representation. The Wigner transform  of the density matrix is defined as
\bea
\label{eq14}
W(q,p,t)=&& \frac{1}{2\pi\hbar}\int_{-\infty}^\infty d\Delta \exp\Big[-\frac{ip\Delta}{\hbar}\Big] \nonumber \\
&& \times\rho\Big(q+\frac{\Delta}{2}, q-\frac{\Delta}{2},t\Big)  \ .
\eea
We mention that  the (pseudo-) phase space distribution $W(q,p,t)$ is also related to the $(k,\Delta)$ representation by a  (double) Fourier transform \cite{giu96}.
Applying the  Wigner transform to the master equation (\ref{eq2}) (retaining again only the last term) we arrive at  a fractional diffusion equation in momentum,
\be
\label{eq15}
\frac{\partial}{\partial t} W(q,p,t) = D_\alpha \, \frac{\partial^\alpha
  }{\partial |p|^\alpha}W(q,p,t) \ .
\ee
Here  the Riesz fractional derivative  is defined
through its Fourier transform as \cite{sam93,sai97}, 
\be 
\label{eq}
 -\frac{\partial^\alpha}{\partial
  |p|^\alpha} = \frac{1}{2\pi}\int_{-\infty}^\infty dz\,
\exp[-ipz]\, |z|^\alpha\ .
\ee
The solution of the fractional equation (\ref{eq15}) can for instance be found in \cite{jes99}.
Next we consider an initial state of our system that consists of the linear superposition of two Gaussian wave packets  separated by a distance $2a$: $\psi(x) = c_1\exp[-(x-a)^2/4b^2]+c_2\exp[-(x+a)^2/4b^2]$. The Wigner function corresponding to this initial state can be written as a sum of two classical contributions plus a quantum (interference) term: $W=W_1+W_2+W_{int}$. For the value  $\alpha\!=\!1$, analytical evaluation of the Wigner function $W(q,p,t)$  is possible. This leads to 
\bea
W_{1,2}(q,p,t) &=&  |c_{1,2}|^2 \exp\left[-\frac{(q \pm a)^2}{2b^2}\right] f(0) \ ,\\
W_{int}(q,p,t) &=& \exp\left[\frac{a^2-q^2}{2b^2}\right] \nonumber \\
&&\times \{c_1^*c_2 f(a)+c_1c_2^*f(-a)\} \ . 
\eea
The function $f(a)$ is given by the expression
\bea
f(a) &=& \sqrt{\frac{b^2}{2\pi}}\left\{ \exp\left[\left(D_\alpha t-
\left(\frac{a}{2b}-ip\right)\right)^2 2b^2\right]\right. \nonumber \\
&&\left.\times\left[ 1-\mbox{Erf}\left(\left(D_\alpha t-\left(\frac{a}{2b}-ip\right)\right) \sqrt{2}b\right)\right]\right. \nonumber \\
&+&\left.\exp\left[\left(D_\alpha t+
\left(\frac{a}{2b}-ip\right)\right)^2 2b^2\right]\right.\nonumber \\  
&&\left.\times\left[ 1-\mbox{Erf}\left(\left(D_\alpha t+\left(\frac{a}{2b}-ip\right)\right) \sqrt{2}b\right)\right]\right\} \ ,
\eea
$\mbox{Erf}(x) = 2/\sqrt{\pi} \int_0^x \exp[-t^2] dx$ being the Error function. Moreover,  in the limit where  the two wave packets are well separated ($a \gg b$), the time dependence of the interference term can be  simplified to become
\be
W_{int}(q,0,t) \sim \sqrt{\frac{b^2}{2\pi}}\exp\left[\frac{a^2}{2b^2}-2aD_1 t\right] \ .
\ee 
Hence the interference between the two superposed Gaussian wave packets is exponentially destroyed  by decoherence on a time $\tau^1_D= [(\Gamma^\downarrow/\hbar) 2a/x_0] ^{-1}$, where $2a$ is the distance between the two wave packets. 

\noindent
{\bf Discussion.} Let us now discuss the results that we have obtained so far. We have found that for a quantum L\'evy process with exponent $\alpha$, the suppression of coherence over a distance $\Delta x$ takes place exponentially on a time scale  $\tau^\alpha_D\sim [(\Gamma^\downarrow/\hbar)(\Delta x/x_0)^\alpha] ^{-1}$, $\Delta x$ being either the width of the single wave packet in the computation  of the linear entropy or  the separation of the two packets in the evaluation of the Wigner function. 
The decoherence time  $\tau^\alpha_D$  is hence inversely proportional to the spreading width $\Gamma^\downarrow$ and also inversely proportional to the $\alpha$th power of the ratio  of the distance $\Delta x$ over  the environment correlation length $x_0$. When compared to the Gaussian case $\alpha=2$, we  find that $\tau^\alpha_D/ \tau^2_D \sim (\Delta x /x_0)^{2-\alpha}$, independent of the factor  $\Gamma^\downarrow/\hbar$. 
\noindent
 In the present analysis the environment correlation length $x_0$ is  much larger than any typical system separation $\Delta x$, so that  we have $\Delta x /x_0 \ll1$. Since  for L\'evy stable motion $\alpha<2$, we are therefore lead to the conclusion that L\'evy decoherence is {\it always} (for any exponent $\alpha$) faster than Gaussian decoherence (or in other words its decoherence time is always  smaller). 
The dependence  of the decoherence time on the value of the stability index is shown in Fig.~(\ref{fig1}), where the linear entropy $S_\alpha(t)$ is  plotted as a function of time for different values of the parameter $\alpha$.  We see that  the smaller the value of $\alpha$, the faster the decoherence for a L\'evy process.  For instance, for  a ratio $\Delta x/x_0= 10^{-3}$,  L\'evy decoherence is three orders of magnitude faster than Gaussian decoherence when $\alpha =1$. L\'evy  anomalous diffusion  hence appears, in this sense, to be  ``more classical'' than its Gaussian counterpart.  

This result can be interpreted as follows. In the Langevin picture of Brownian motion, the coupling to the environment manifests itself in the addition of  two forces to Newton's equation: first, a friction force which dissipates energy away from the system and, second, a fluctuating stochastic force which randomly injects energy back into the system.  For  L\'evy stable motion, the variance of this stochastic force is divergent which means that the environment randomly supplies an infinite amount of energy to the system (this is, by the way, the physical origin of the L\'evy flights) \cite{wes82}. Thus, in this case,  the coupling to the  surroundings  has a much stronger influence on the evolution of the system and, in turn, decoherence, like diffusion, is greatly enhanced.  Note, however, that in contrast to diffusion, L\'evy decoherence does not involve divergent quantities.

In summary, we have investigated the decoherence properties of a small quantum system weakly coupled to a chaotic background when its dynamics exhibits L\'evy type anomalous behavior. We have determined the decoherence time, first, by computing the linear entropy of the system and, then, by examining the decay of interference fringes of two superposed Gaussian wave packets in the Wigner representation. In comparison  to the  Gaussian case, we have found that L\'evy decoherence is {\it always} faster. More precisely, the deviation of the decoherence time from its Gaussian value is larger the smaller the value of the characteristic exponent $\alpha$.

\noindent
This work was supported by the RTD network Nanoscale Dynamics, Coherence and Computation. We thank D. Kusnezov and C. Lewenkopf for discussion. 


\begin{references}
\bibitem{rem1} It should be noted that the limit $x_0\rightarrow \infty$ also corresponds to the weak--coupling limit (see Ref.~\cite{kus99}).
\end{references}
\end{document}